\documentclass[fleqn,twoside]{article}
\usepackage{espcrc2}

\usepackage{graphicx}
\usepackage{epsfig}
\usepackage[figuresright]{rotating}

\newcommand{\AmS}{{\protect\the\textfont2
  A\kern-.1667em\lower.5ex\hbox{M}\kern-.125emS}}

\title{Generalized parton distributions of the pion}

\author{F. Bissey\address[MCSD]{Universit\'e de Li\`ege, D\'epartement de Physique B5, \\ 
        B-4000 Sart Tilman Li\`ege 1, Belgium}%
        \thanks{Present address: Institute of Fundamental Sciences, Massey University, Private Bag 11~222, Palmerston North 5301, New Zealand.},
        J. R. Cudell\addressmark,
        J. Cugnon\addressmark\thanks{e-mail: J.Cugnon@ulg.ac.be}, J. P. Lansberg\addressmark, P. Stassart\addressmark\thanks{This work was supported by the European Union contract N$^{\circ}$ HPRN-CT-2000-00130 (ESOP collaboration)}
}
       
\begin{document}

\begin{abstract}
Off-forward structure functions of the pion are investigated in twist-two and twist-three approximation. A simple model is used for the  pion, which allows to introduce finite size effects, while preserving gauge invariance. Results for the imaginary parts of the $\gamma^* \pi \rightarrow \gamma^* \pi$ off-forward amplitude and  of the structure functions are presented. Generalized Callan-Gross relations are obtained.
\vspace{1pc}
\end{abstract}

\maketitle

\section{INTRODUCTION}
Structure functions are  useful tools to understand the structure of hadrons. At large $Q^2$, they are related to  parton distributions. Although their $Q^2$-evolution  is consistent with perturbative QCD, their bulk properties come from nonperturbative effects. The latter are often treated by low-energy models, such as NJL, which establish a connection whith the low $Q^2$ physics. An extensive work has been done during the last years along these lines~\cite{diagonal}. 

The interest has turned to off-diagonal structure functions. The latter appear as convolutions of generalized parton distributions, which, in some way, carry information about correlations between partons. They can be related to the off-forward $\gamma^*$-hadron amplitude. In order to illustrate the properties of these quantities, we undertook to calculate them in the simple case of  the pion. In Ref.~\cite{ours}, we first calculated the forward amplitude and the quark distribution in a simple model. We considered that the pion field is coupled to (constituent) quark fields through a simple $\gamma^5$ vertex. Furthermore, we introduced the effects of the pion size through a gauge-invariant procedure by requiring that the squared relative momentum of the quarks inside the pion is smaller than a cut-off value. The most remarkable result of this investigation is that the momentum fraction carried   by the quarks is  smaller than  one, although gluonic degrees of freedom are not included. Here, we report on the extension of our model to the off-diagonal case.
     
\section{TENSORIAL STRUCTURE OF THE   $\gamma^{\star} \pi \rightarrow 
\gamma^{\star}\pi$ AMPLITUDE }

\begin{figure*}[t]
\centering
\quad\quad\mbox{{\includegraphics[height=3.0cm]{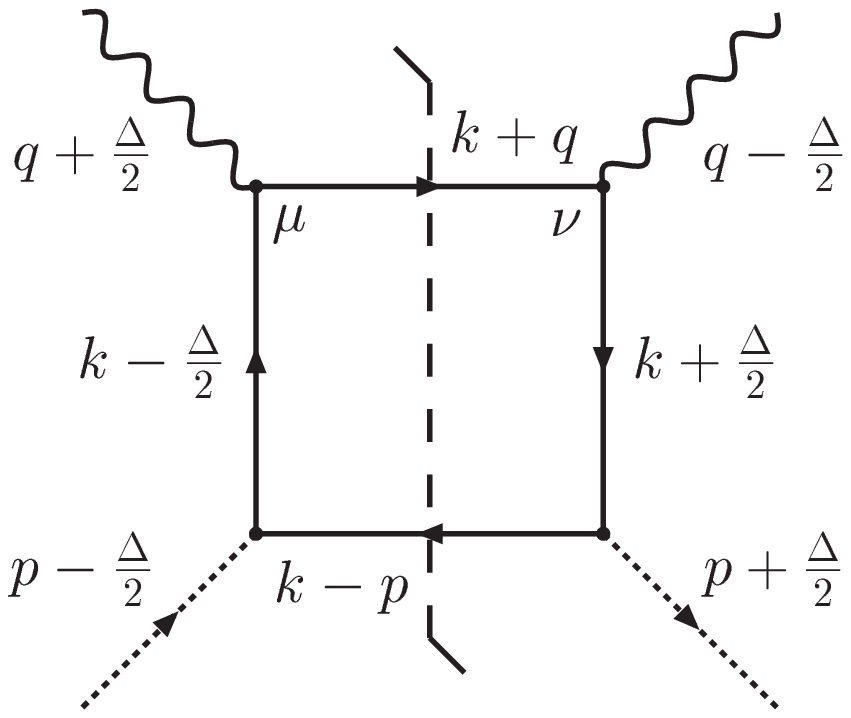}}\quad
      {\includegraphics[height=3.0cm]{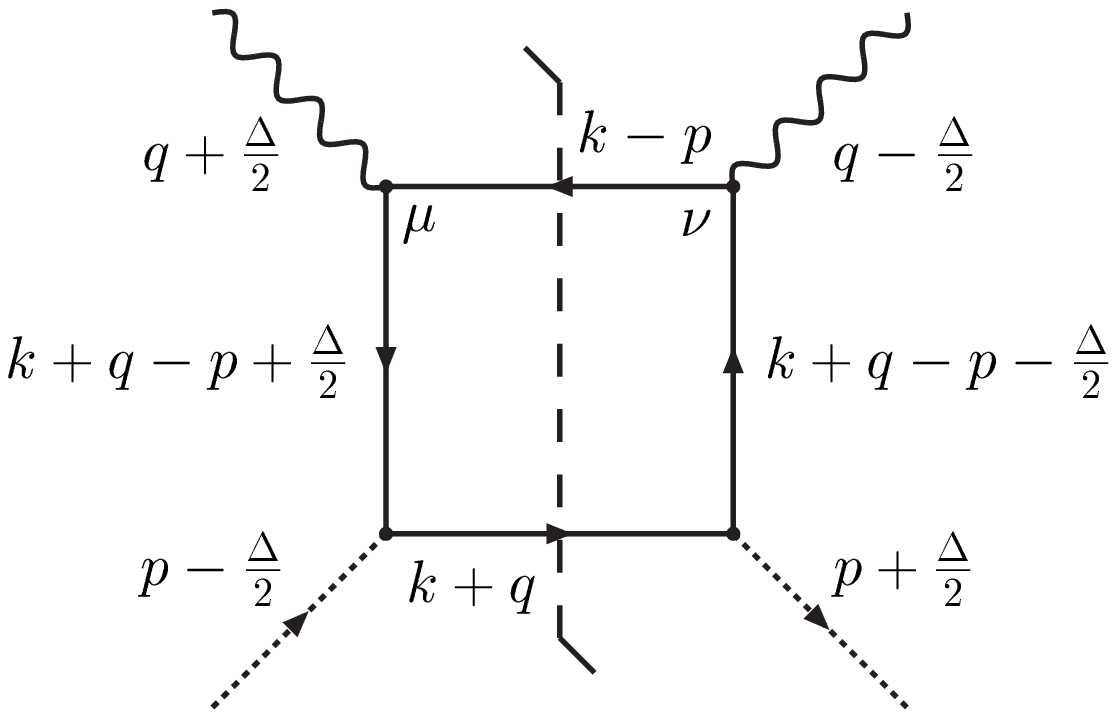}}}
\raisebox{-4cm}{{\includegraphics[height=3.0cm]{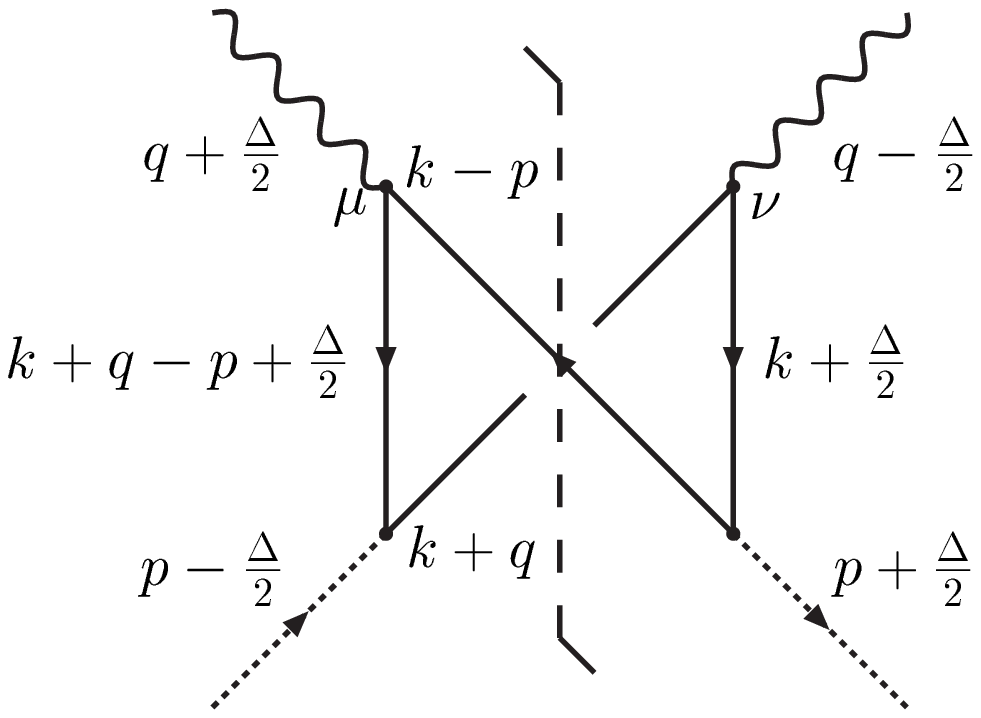}}\quad\quad\quad
      {\includegraphics[height=3.0cm]{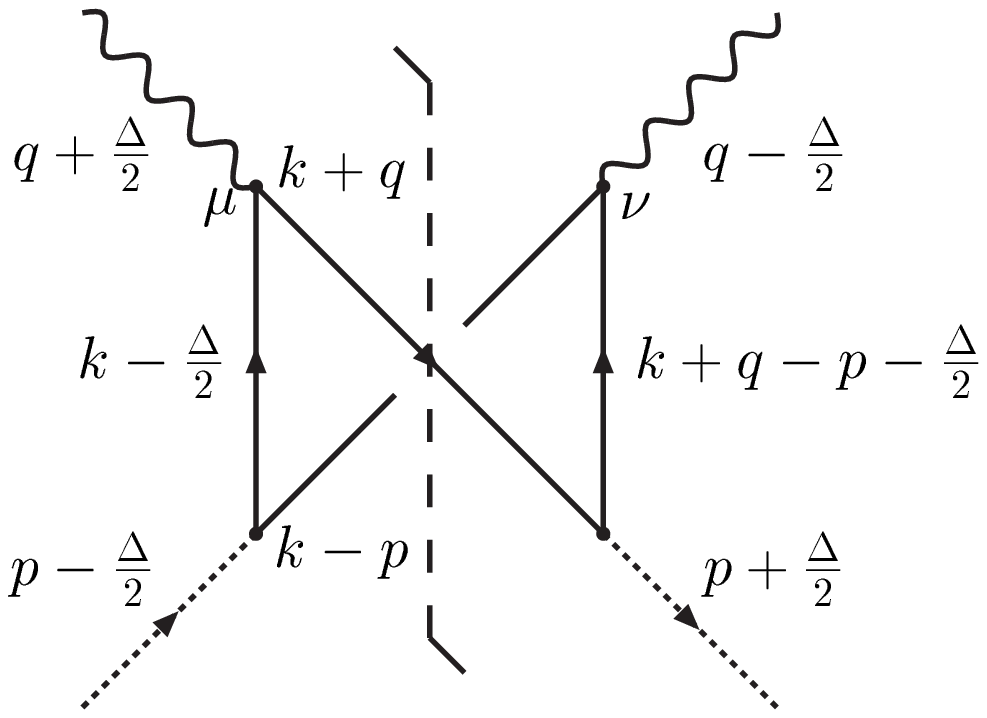}}}
\caption{Simplest diagrams contributing to the imaginary part of the 
amplitude for   $\gamma^{\star} \pi \rightarrow 
\gamma^{\star}\pi$ scattering. Upper
(lower) diagrams are referred to  as box (crossed) diagrams. Dashed lines
represent the discontinuity of the amplitudes or their imaginary part.}
\label{f1}
\end{figure*}

We adopt the kinematics shown in Fig.~\ref{f1}. We will use the  Lorentz invariants $t = \Delta^2$,  $Q^2 = -q^2$, 
$x = Q^2/2p\cdot q$ and $\xi = \Delta\cdot q / 2p\cdot q$. The diagonal limit is characterised by $\xi=t=0$ and the elastic limit by $\xi=0$. 

The  hadronic tensor $T_{\mu\nu} (q, p, \Delta)$ can be written, for a scalar or pseudoscalar target, as~\cite{Bel}
\begin{eqnarray}
T_{\mu\nu}(q, p, \Delta) =
- {\cal P}_{\mu\sigma} g^{\sigma\tau} {\cal P}_{\tau\nu}
F_1
+ \frac{{\cal P}_{\mu\sigma} p^\sigma p^\tau {\cal P}_{\tau\nu}}{p \cdot q}
F_2\nonumber\\
+ \frac{{\cal P}_{\mu\sigma} (p^\sigma (\Delta^\tau-2\xi p^\tau)
+  (\Delta^\sigma-2\xi p^\sigma) p^\tau) {\cal P}_{\tau\nu}}{2 p \cdot q}
F_3\nonumber\\
+ \frac{{\cal P}_{\mu\sigma}
(p^\sigma (\Delta^\tau-2\xi p^\tau)
-  (\Delta^\sigma-2\xi p^\sigma)p^\tau) {\cal P}_{\tau\nu}}{2 p \cdot q}
F_4\nonumber\\
\hspace{-1cm}+ {\cal P}_{\mu\sigma}
(\Delta^\sigma-2\xi p^\sigma) (\Delta^\tau-2\xi p^\tau) {\cal P}_{\tau\nu}
F_5.
\end{eqnarray}
Current conservation is guarenteed by means of the projector
\begin{equation} 
{\cal P}_{\mu\nu}=g_{\mu\nu} - q_{2 \mu} q_{1 \nu}/ q_1 \cdot q_2,
\end{equation}
where $q_1$ and $q_2$ are the momenta of the incoming and outgoing photons, respectively. The structure functions $F_i$ are functions of the invariant quantities $x$, $\xi$ and $t$. They are all even functions of $\xi$, except for $F_3$, which is odd.

\section{THE MODEL}

The  model introduced in our previous work~\cite{ours} includes massive pion and massive quark
fields and a pion-quark coupling described by the following Lagrangian  interaction density
\begin{equation} 
 {\mathcal L}_{int}= i g (\overline{\psi}\mbox{  $\vec{\tau}$} \gamma_5 \psi) .
\mbox{  $\vec{\pi}$},
\end{equation}
where $\psi$ is the quark field, $\vec{\pi}=(\pi^+,\pi^0,\pi^-)$ is the pion field and $\vec{\tau}$ is the isospin  operator.

At leading order in the loop expansion, four diagrams  contribute. They 
are displayed in Fig.~\ref{f1}. We have evaluated their imaginary part, using the integration variables $\tau=k^2$, $k_{\rho}=|\vec{k}|$, $\phi$ and $\theta$,  the polar angles of $\vec{k}$ with respect
to  the direction of the incoming photon.   Actually, due to the discontinuity of the diagrams, indicated in Fig.~\ref{f1}, the integration bears on  $\tau$ and $\phi$ only. We do not give  the  expressions here. They can be found in Ref.~\cite{our2} . However, we can sketch our procedure for imposing a finite size to the pion. The relative four-momentum squared of the quarks inside the pion is given by

\begin{eqnarray}
O^\pm&=&\left(2k-p\pm\frac{\Delta}{2}\right)^2 \\&=&
2\tau + 2 m_q^2 - m_{\pi}^2+\frac{t}{2}\pm 2k\cdot \Delta,
\end{eqnarray}
for pion-quark vertices like the ones in the first diagram of Fig.~\ref{f1}. Note that
this can be rewritten as 
\begin{equation}
O^\pm=\left(k \pm \frac{\Delta}{2}\right)^2 +  2 m_q^2 - m_{\pi}^2.
\end{equation}
The first quantity in the r.h.s. being nothing but  the squared momentum transfer for the $\gamma^{\star} \pi \rightarrow q \overline{q}$ process, $O^\pm$ can be written as function of the external variables for this process and of the masses. Similar expressions hold for other vertices. Generalizing the procedure of Ref.~\cite{ours}, we  require $|O^{\pm}| <
\Lambda^2$ either for one or  the other vertex of each diagram. Gauge invariance is therefore preserved by this cut-off, as it can be thought of as a constraint on the intermediate state cut lines. In practice, this is equivalent to requiring one of the two following conditions:
\begin{equation}
\begin{array}{l}
\tau >-\frac{\Lambda^2}{2}+ \frac{m_{\pi}^2}{2}-m_q^2-\frac{t}{4}+\left|k\cdot \Delta\right|,\\
\\
\tau < \frac{\Lambda^2}{2}- \frac{m_{\pi}^2}{2}+3m_q^2+\frac{t}{4}-\frac{Q^2}{x}
-\left|\frac{\xi Q^2}{x}+k\cdot \Delta\right|.
\end{array}
\end{equation}
As explained in Ref.~\cite{ours}, owing to these conditions and for small $t$, the crossed diagrams
 are  suppressed
by a power $\Lambda^2/Q^2$, compared to  the box diagrams.

We keep the  coupling constant $g$ as in the diagonal case, where it was  determined by imposing that there are only two constituent quarks in the pion, or equivalently that the following relation
\begin{equation}
\int _{0}^{1}F_{1}dx=\frac{5}{18}
\end{equation}
holds, which makes $g$ dependent upon $Q^2$. It turns out that, with the cut-off, $g$ reaches an asymptotic value for $Q^2$ above 2 GeV$^2$. We will use this value below. 
\section{RESULTS}
We investigated the main properties of the structure functions $F_i$ in many different domains of the parameters. We cannot give an overview of the results here. We will instead concentrate on two particular results, namely the behaviour at high $Q^2$ and the momentum sum rule. The results presented below pertain to the chiral limit ($m_{\pi}$=0). They have been obtained with the same cut-off value as in Ref.~\cite{ours}, namely $\Lambda$= 0.75 GeV, a value consistent with the hard core radius of the pion in the chiral bag model~\cite{BR86}.

\subsection{High $Q^2$ limit:  generalised Callan-Gross relations}

Expanding the ratios  
$\frac{F_2}{F_1}$,$\frac{F_3}{F_1}$,$\frac{F_4}{F_1}$,$\frac{F_5}{F_1}$, 
we obtain the following asymptotic behaviours:
\begin{eqnarray}
F_2&=&2xF_1+{\cal O}(1/Q^2), \label{eq:GCG1}    \\
F_3&=&\frac{2x\xi}{\xi^2-1}F_1+{\cal O}(1/Q^2)  \label{eq:GCG2}    \\
F_4&=&\frac{2x}{\xi^2-1}F_1+{\cal O}(1/Q^2)  \label{eq:GCG3}         \\
F_5&=&{\cal O}(1/Q^2),\label{eq:GCG4} 
\end{eqnarray}
The first relation is similar (at leading order in $1/Q^2$ and with the replacement of $x$ by $x_B$) to the Callan-Gross relation 
between the diagonal structure functions $F_1$ and $F_2$. Except for $F_5$, which is small at large $Q^2$, these relations 
show that $F_2$, $F_3$ and $F_4$ are simply related to $F_1$ at leading order. They are consistent with  the 
parity of these functions with respect to $\xi$. 
They constitute a remarkable
result of our model. Furthermore, we checked that the term ${\cal O}(1/Q^2)$
in Eq.~(\ref{eq:GCG1}) is numerically quite small, even for moderate $Q^2$. One may wonder whether these properties are 
typical of our model  or more general.

\subsection{The momentum sum rule}

We recall that, for the diagonal case (elastic scattering in the forward direction), we found in Ref.~\cite{ours} that the momentum fraction carried by the quarks
\begin{equation}
2<x>=\frac{\int_{0}^{1}2x F_{1}dx}{\int_{0}^{1}F_{1}dx}
\end{equation}
has a rather low value, $\approx$0.6, independently of $Q^2$ (as long as it is larger than $\sim$
2 GeV$^2$). The quantity  2$<x>$ moves slowly to unity when the cut-off is removed, showing that the low value mentioned above is due to finite size effects.

\begin{figure}[h]
\centering
\includegraphics[height=5.0cm]{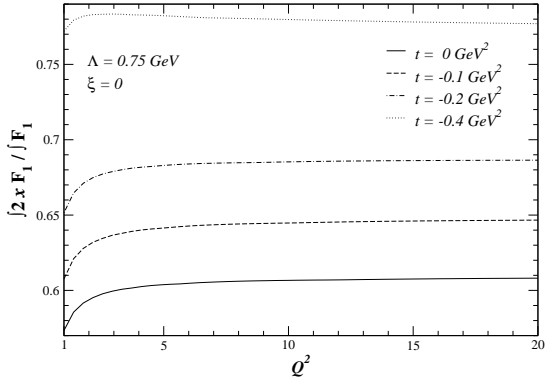}
\caption{Mean value of 2$x$ with respect to the $F_1$ distribution, as function of $Q^2$, for various values of $t$, in the elastic case. }
\label{f2}
\end{figure}

We investigated how this property evolves when going to the off-forward case. Some results are given in Fig.~\ref{f2} for the elastic case ($\xi$=0). One can see that 2$<x>$ increases when $|t|$ increases:  the quantity which measures the momentum fraction carried by the quarks in the diagonal case and probed by the process increases with the momentum transfer. Strictly speaking $F_{1}$ is not the quark momentum distribution for $\xi \neq$0. By continuity, it is expected to keep this property for small values of this variable. This point however deserves further investigation. 

\subsection{Other results}
We have also obtained many other results. It is not the place to discuss them in detail. Let us just mention that we analysed  the twist decomposition of the $F_i$'s and worked out the twist-2 $\cal H$ and twist-3 ${\cal H}^3$, $\tilde{\cal H}^3$ contributions, as defined in Ref.~\cite{Bel}. We found that the twist-3 contributions are not suppressed at large $Q^2$. This is a remarkable and somewhat unexpected result of our model.

\section{CONCLUSION}
We have extended our previous model for the pion to investigate  the off-diagonal structure 
functions. We recall the two main results of our previous investigation.
First, the introduction of a cut-off forces the crossed diagrams to behave
as higher-twists and to relate the imaginary part of the forward amplitude with quark
distributions. Second, the pion size effects, embodied by the cut-off, do not allow the fullfillment of 
 the momentum sum rule ($2 \left< x \right>=1$) at infinite $Q^2$. Physically, this corresponds
to the fact that the quarks can never be considered as free. 

We found that these characteristics are qualitatively preserved
when going off-forward, at least for small or moderate excursions off the forward
case. Our results rest on the analysis of the imaginary part of the amplitude only. However, in this regime at least, the introduction of the real part, which can be obtained via dispersion relations (up to a subtaction constant), is not expected to bring drastic changes.

Our investigation yields interesting and somehow unexpected results. In particular, 
we singled out generalized Callan-Gross relations (Eqs.9-11), which link the $F_i$'s in a simple manner at leading
order in $1/Q^2$. More intriguing are the twist-three structure functions, that  do not scale as  a power of  $1/Q$. At this point of our investigation, we are not able to state whether these properties are typical
of our model or are more general. In particular, it would be interesting to know how these results 
are affected when turning to the full amplitude.

\end{document}